\documentclass[12pt]{article}

\usepackage{amsmath,a4wide,amssymb,amsthm}
\usepackage{epsfig}

\newtheorem{theorem}{Theorem}   
\newtheorem{lemma}[theorem]{Lemma}

\numberwithin{equation}{section}
\numberwithin{theorem}{section}

\title{\bf Staircase polygons: moments of diagonal lengths and column heights}

\author{\sc Christoph~Richard\\
\\
{\it Fakult\"at f\"ur Mathematik, Universit\"at Bielefeld,}\\
{\it Postfach 10 01 31, 33501 Bielefeld, Germany}}        

\begin{document}

\maketitle

\begin{abstract}
We consider staircase polygons, counted by perimeter and sums of 
$k$-th powers of their diagonal lengths, $k$ being a positive integer.
We derive limit distributions for these parameters in the limit of 
large perimeter and compare the results to Monte-Carlo simulations 
of self-avoiding polygons. We also analyse staircase polygons, counted
by width and sums of powers of their column heights, and we apply our methods 
to related models of directed walks.
\end{abstract}

\centerline{\it dedicated to Tony Guttmann on the occasion of 
his 60th birthday}

\section{Introduction}

Self-avoiding polygons are (images of) simple closed curves on the square 
lattice. We are interested in counting such polygons by perimeter, enclosed 
area, and other quantities, where we identify polygons which are equal up to 
a translation.  In statistical physics, self-avoiding polygons, counted 
by perimeter and area, serve as a model of two-dimensional vesicles, 
where the vesicle pressure is conjugate to the polygon area \cite{BMS91,FGW}. 
Despite the long history of the problem (see \cite{MS93,J00} for reviews), there 
is not much exact, let alone rigorous knowledge. Recent progress comes from 
two different directions. It appears that stochastic methods, 
combined with ideas of conformal invariance, describe a number of fractal 
curves appearing in various models of statistical physics, including 
self-avoiding polygons \cite{LSW02}. This will not further be discussed 
in this article. 

Another line of research consists in analysing solvable subclasses of 
self-avoiding polygons. Such classes of polygons have been studied for 
a long time and by different people, see e.g.~\cite{BM96} and references 
therein for (subclasses of) column-convex polygons. In this article, we 
will concentrate on one such subclass, the class of staircase 
polygons. The perimeter generating function of staircase polygons has 
been given by Levine \cite{L59} and by P\'olya \cite{P69}, the width 
and area generating function has been given by Klarner and Rivest 
\cite{KR74} and by Bender \cite{Be74}, the perimeter and area generating 
function of this model has been given by Brak and Guttmann \cite{BG90}, 
Lin and Tseng \cite{LT91}, Bousquet-M\'elou and Viennot \cite{BMV92}, 
Prellberg and Brak \cite{PB95}, and by Bousquet-M\'elou \cite{BM96},
using various solution methods.

Of particular interest to the Melbourne group of statistical mechanics 
and combinatorics is the critical behaviour of such models, in particular 
their critical exponents and scaling functions. For staircase polygons, 
counted by perimeter and area, these quantities had been obtained in 1995 
\cite{P95,PB95}.
On the basis of the numerical observation that rooted self-avoiding 
polygons, counted by perimeter and area, have the same critical 
exponents as staircase polygons, Tony Guttmann raised the question 
\cite{PO95b} whether their scaling functions, which describe the 
vesicle collapse transition, might be the same. One of the 
implications of this conjecture concerns the asymptotic distribution 
of polygon area in a uniform model, where polygons of fixed perimeter 
occur with equal weight. This distribution was tested for by self-avoiding 
polygon data from exact enumeration and found to be satisfied to a 
high degree of numerical accuracy \cite{RGJ01}. A field-theoretic 
justification for the form of the scaling function was then given by Cardy 
\cite{C01}. Later, other implications of the scaling function conjecture 
relating to the asymptotic distribution of perimeter at criticality and to 
the crossover behaviour were found to be satisfied, within numerical 
accuracy \cite{RJG04}.

\begin{figure}[htb]
\begin{center}
\begin{minipage}[b]{0.95\textwidth}
\center{\epsfig{file=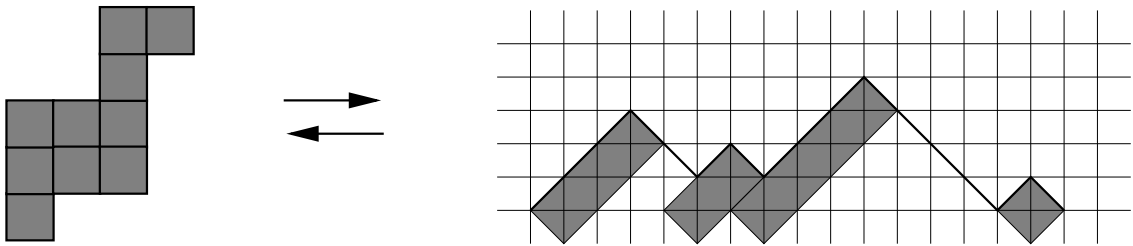,width=15cm}}
\end{minipage}
\end{center}
\caption{\label{fig:pol}
\small The set of staircase polygons is in one-to-one correspondence with 
the set of Dyck paths \cite{DV84,S99}. Vertical column 
heights of a polygon correspond to peak heights of a path.}
\end{figure}

\begin{figure}[htb]
\begin{center}
\begin{minipage}[b]{0.95\textwidth}
\center{\epsfig{file=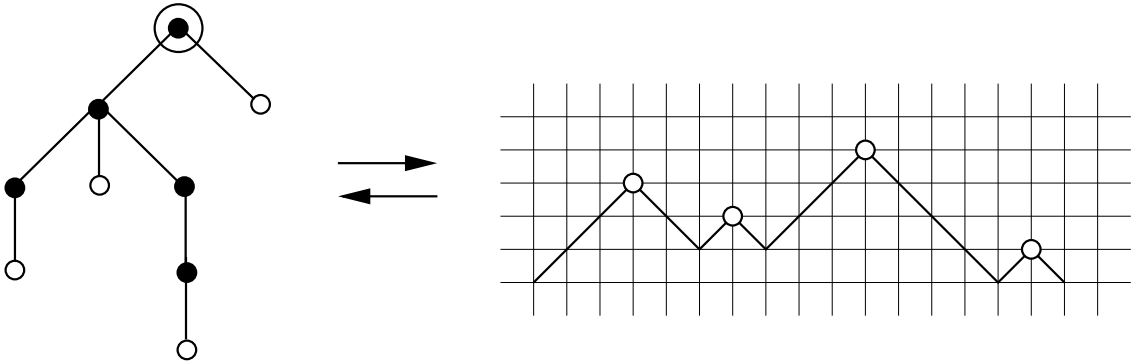,width=15cm}}
\end{minipage}
\end{center}
\caption{\label{fig:tree}
\small The set of rooted, ordered trees is in one-to-one correspondence with
the set of Dyck paths \cite{DV84,S99}. Leave distances to the root of a tree 
correspond to peak heights of a path.}
\end{figure}

One might thus be tempted to conclude that critical self-avoiding polygons 
and critical staircase polygons lie in the same universality class. Then, 
one should expect that also distributions of parameters different from the 
area should coincide for both models. Investigating the implications of this 
idea is one of the aims of this article. We thus re-analyse staircase 
polygons, which we count by a number of parameters generalising the area. 
In so doing, we adopt the viewpoint of analytic combinatorics \cite{SF05}, 
thereby highlighting connections to statistical physics and to a 
probabilistic description. For example, staircase polygons are in 
one-to-one correspondence to Dyck paths \cite{DV84,S99}, see Figure 
\ref{fig:pol} for an illustration and below for definitions.
In turn, Dyck paths are classical objects of enumerative combinatorics
\cite{S99}, which are asymptotically described by Brownian excursions. 
This connection is described at various places in the literature, see e.g. 
\cite{K76,T91,A92}. Moreover, Dyck paths are closely related to models of trees 
\cite{DV84,S99}, as indicated in Figure \ref{fig:tree}.
In fact, the generating functions of polygon models, models of (simply
generated) trees, and models of paths satisfy similar functional equations, 
typically leading to limit distributions of the same type. More
generally, it appears that limit distributions for counting parameters 
are irrelevant of many details of the functional equation. This phenomenon 
of universality has recently been investigated using the setup of 
$q$-functional equations \cite{R05} and extends previous results \cite{D99, R02} 
by including counting parameters which generalise the polygon area. The 
results in \cite{R05} were obtained by an interplay 
between methods from statistical physics, combinatorics and stochastics. 
Corresponding counting parameters appeared previously within a 
combinatorial setup, see Duchon \cite{D99} and Nguy$\tilde{\mbox{\rm \^e}}$n 
Th$\acute{\mbox{\rm \^e}}$ \cite{NT03, NT03b}. The derivation of moment 
recurrences of limit distributions for these parameters is inspired 
by the method of dominant balance as used by Prellberg and Brak \cite{PB95}
in polygon statistical physics, and some results for the 
corresponding stochastic objects appear to be unknown in the 
stochastics literature.

The outline of this article is as follows. We first derive a functional
equation for the generating function of staircase polygons, counted 
by perimeter and sums of  $k$-th powers of lengths of diagonals, 
$k$ being a positive integer. We also analyse a corresponding quantity
for column heights. We then derive limit distributions for the above
counting parameters in the limit of large perimeter, resp.~width. We will
first review the case $k=1$, which serves as a preparation for the
following section, where the case of general $k$ is discussed. We then investigate the question 
of universality of our results by comparing them to Monte-Carlo simulations 
of self-avoiding polygons. The simulations indicate that the considered parameters 
have different laws for self-avoiding polygons and for staircase polygons. 
The methods discussed in this paper can be applied to similar problems concerning
models of walks and models of trees. This we indicate for 
various models of discrete walks related to Dyck paths, such as meanders and
bridges. The analysis extends previous work \cite{NT03,NT03b}, 
where mainly the case $k=2$ is studied. 

\section{Functional equations}\label{sec:feq} 

\begin{figure}[htb]
\begin{center}
\begin{minipage}[b]{0.95\textwidth}
\center{\epsfig{file=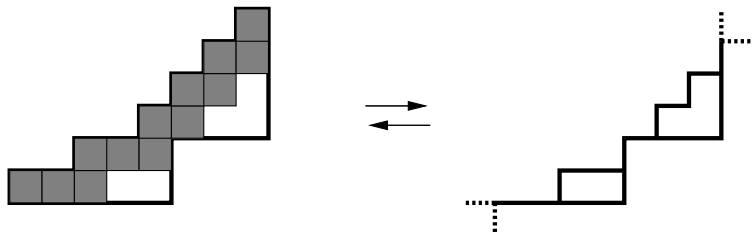,width=10cm}}
\end{minipage}
\end{center}
\caption{\label{fig:rem}
\small The set of staircase polygons is in one-to-one correspondence with 
the set of ordered sequences of edges and staircase polygons. A corresponding
bijection is geometrically characterised by shifting the upper walk of a 
staircase polygon one unit down and one unit to the left, and by then removing
the first and the last edge of each walk.}
\end{figure}

Consider two fully directed walks on the edges of the square lattice (i.e., 
walks stepping only up or right), which both start at the origin and end 
in the same vertex, but have no other vertex in common. The edge set of such 
a configuration is called a {\em staircase polygon}, if it is nonempty. Each 
staircase polygon is a self-avoiding polygon. For a given staircase polygon, 
consider the construction of moving the upper directed walk one unit down and 
one unit to the right. For each walk, remove its first and its last edge.
The resulting object is a sequence of (horizontal and vertical) edges and 
staircase polygons, see Figure \ref{fig:rem}. The unit square yields the
empty sequence.
Using this construction, it is easy to see that the set of 
staircase polygons $\cal P$ is in one-to-one correspondence with the 
set $\cal Q$ of ordered sequences of edges and staircase polygons. Let us 
denote the corresponding map by $f:\cal P\to \cal Q$. Thus, for a staircase 
polygon $p\in\cal P$, we have $f(p)=(q_1,\ldots,q_n)\in\cal Q$, where $q_i$ is, 
for $i=1,\ldots, n$, either a single edge or a staircase polygon. The image of 
the unit square is the empty sequence $n=0$. A variant of this decomposition 
will be used below in order to derive a functional equation for the 
generating function of staircase polygons.

\begin{figure}[htb]
\begin{center}
\begin{minipage}[b]{0.95\textwidth}
\center{\epsfig{file=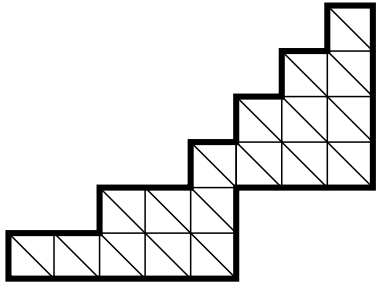,width=5cm}}
\end{minipage}
\end{center}
\caption{\label{fig:diag}
\small A staircase polygon of half-perimeter 14 and area 18. It has 8 diagonals of
length one and 5 diagonals of length two.}
\end{figure}

The horizontal (vertical) perimeter of a staircase polygon is the number 
of its horizontal (vertical) edges, its (total) perimeter is the sum of 
horizontal and vertical perimeter. The horizontal half-perimeter is also called
the width of the polygon, the vertical half-perimeter is also called the height
of the polygon. The area of a staircase polygon is the number 
of its enclosed squares. The half-perimeter of a staircase polygon is 
equal to the number of its (negative) diagonals plus one, 
the area of a staircase polygon equals the sum of the lengths of its 
(negative) diagonals, see Figure \ref{fig:diag} for an illustration.
For $k$ a nonnegative integer, we will consider {\it $k$-th diagonal length 
moments} of a staircase polygon $p$, defined by
\begin{equation}\label{form:dm}
n_k(p)=\sum_{d\in D(p)} l(d)^k,
\end{equation}
where the summation ranges over the set of diagonals $D(p)$ of $p$,  
with $l(d)$ denoting the length of $d$, i.e., the number of squares 
crossed by $d$. We count staircase polygons by 
half-perimeter $n_0(p)+1$ and by the parameters $n_k(p)$, where $k=1,\ldots, M$. 
Let $\boldsymbol u= (u_0,\ldots, u_M)$ denote formal variables. For a staircase 
polygon $p$, define its weight $w_p(\boldsymbol u)$ by
\begin{equation*}
w_p(\boldsymbol u)=u_0^{n_0(p)+1}u_1^{n_1(p)}\cdot\ldots\cdot u_M^{n_M(p)},
\end{equation*}
and let $G(\boldsymbol u)=\sum_{p\in\cal P} w_p(\boldsymbol u)$ denote the 
generating function of $\cal P$. The bijection described above
can be used to derive a functional equation for $G(\boldsymbol u)$.

\begin{theorem}\label{theo:feq}
The generating function $G(\boldsymbol u)$ of staircase polygons satisfies
\begin{equation}\label{form:fe}
G(\boldsymbol u)=\frac{u_0^2 u_1\cdot\ldots\cdot u_M}
{1-2u_0\cdot\ldots\cdot u_M-G(\boldsymbol v(\boldsymbol u))},
\end{equation}
where the functions $v_k({\boldsymbol u})$ are given by
\begin{equation}\label{form:vk}
v_k({\boldsymbol u})=\prod_{l=k}^M u_l^{\binom{l}{k}} \qquad (k=0,1,\ldots,M).
\end{equation}
\end{theorem}

\begin{proof}
For a sequence $q=(q_1,\ldots q_n)\in\cal Q$, its preimage 
$p=f^{-1}(q)\in\cal P$ can be described in the following way, which will prove
more convenient for enumeration purposes. Consider the sequence $(p_1,\ldots, 
p_n)$ of staircase polygons, where $p_i=f^{-1}((q_i))$, for $i=1,\ldots,n$. The 
staircase polygon $p$ is obtained from $(p_1,\ldots,p_n)$ by concatenation, i.e., 
by translating $p_{i+1}$ such that the top right square of $p_i$ and the lower 
left square of $p_{i+1}$ coincide, for $i=1,\ldots, n-1$. Let us denote by 
$\widetilde{\cal P}\subset \cal P$ the subset of staircase polygons such that
for each $p\in\widetilde{\cal P}$ we have $p=f^{-1}(q)$, where $q$ is either a single 
edge or a staircase polygon. The above description implies that the set of staircase
polygons is in one-to-one correspondence to the set of concatenated ordered 
sequences of staircase polygons from $\widetilde{\cal P}$.

For a single edge $e$, the weight of the staircase polygon $f^{-1}(e)$ is given by 
$w_{f^{-1}(e)}(\boldsymbol u)=u_0^3u_1^2\cdot\ldots\cdot u_M^2$. For a staircase 
polygon $p$, note that by the binomial theorem we have
\begin{displaymath}
n_k(f^{-1}(p))=\sum_{d \in D(f^{-1}(p))} l(d)^k = 2+\sum_{d\in D(p)} (l(d)+1)^k = 
2+\sum_{i=0}^k \binom{k}{i} n_i(p).
\end{displaymath}
This implies for the weights
\begin{displaymath}
\begin{split}
w_{f^{-1}(p)}(\boldsymbol u) &=  u_0 \prod_{l=0}^M u_l^{n_l(f^{-1}(p))} 
= u_0\prod_{l=0}^M \left(u_l^2 \prod_{k=0}^l u_l^{\binom{l}{k} n_k(p)}\right)
= u_0 \left(\prod_{l=0}^M u_l^2\right) \left(\prod_{k=0}^M 
\prod_{l=k}^M u_l^{\binom{l}{k} n_k(p)} \right)\\
&= u_0^2 \left(\prod_{l=1}^M u_l\right) \left(\prod_{l=0}^M u_l u_l^{n_0(p)}\right) 
\left(\prod_{k=1}^M \prod_{l=k}^M u_l^{\binom{l}{k} n_k(p)} \right)\\
&= u_0^2 \left(\prod_{l=1}^M u_l\right) v_0(\boldsymbol u) 
\prod_{k=0}^M v_k(\boldsymbol u)^{n_k(p)}
= u_0^2 u_1\cdot \ldots\cdot u_M 
w_{p}(\boldsymbol v(\boldsymbol u)),
\end{split}
\end{displaymath}
where the functions $v_k(\boldsymbol u)$ are given by \eqref{form:vk}.
For a sequence $(p_1,\ldots,p_n)$ of staircase polygons, denote their 
concatenation by $c(p_1,\ldots,p_n)$. We have $w_{c(p_1,p_2)}
(\boldsymbol u)=u_0^{-2}u_1^{-1}\cdot\ldots\cdot u_M^{-1}w_{p_1}
(\boldsymbol u)w_{p_2}(\boldsymbol u)$, a monomial 
in $\boldsymbol u$, and thus 
\begin{displaymath}
w_{c(p_1,\ldots,p_n)}
(\boldsymbol u)=\frac{1}{u_0^{2(n-1)}u_1^{n-1}\cdot\ldots\cdot u_M^{n-1}}
w_{p_1}(\boldsymbol u)\cdot\ldots\cdot w_{p_n}(\boldsymbol u).
\end{displaymath}
The above bijection yields a functional equation for the generating function 
$G(\boldsymbol u)$. We have
\begin{displaymath}
\begin{split}
G(\boldsymbol u) &= \sum_{n=0}^\infty \sum_{(p_1,\ldots,p_n)\in
(\widetilde{\cal P})^n} w_{c(p_1,\ldots,p_n)}(\boldsymbol u)\\
&= \sum_{n=0}^\infty u_0^2u_1\cdot\ldots\cdot u_M \sum_{(p_1,
\ldots,p_n)\in(\widetilde{\cal P})^n} \frac{w_{p_1}(\boldsymbol u)}
{u_0^2u_1\cdot\ldots\cdot u_M}\cdot\ldots\cdot\frac{w_{p_n}(
\boldsymbol u)}{u_0^2u_1\cdot\ldots\cdot u_M}\\
&= u_0^2u_1\cdot\ldots\cdot u_M \sum_{n=0}^\infty \left( 
\frac{1}{u_0^2u_1\cdot\ldots\cdot u_M}\sum_{p\in\widetilde{\cal P}} 
w_{p}(\boldsymbol u)\right)^n\\
&= u_0^2u_1\cdot\ldots\cdot u_M  \frac{1}{1-\frac{1}{u_0^2u_1\cdot
\ldots\cdot u_M} \left( 2w_{f^{-1}(e)}(\boldsymbol u) + 
\sum_{p\in\cal P} w_{f^{-1}(p)}(\boldsymbol u)\right)
}\\
&= \frac{u_0^2u_1\cdot\ldots\cdot u_M}{ 1- 2u_0u_1\cdot\ldots
\cdot u_M-G(\boldsymbol v(\boldsymbol u))}
\end{split}
\end{displaymath}
This is the functional equation \eqref{form:fe}.
\end{proof}

\noindent {\bf Remark.} The case $M=0$ is standard, see e.g.~\cite{S99}. 
The case $M=1$ has been studied by various techniques in 
\cite{G80,DV84,PB95,BM96}. For corresponding models of walks and trees, 
see also the review in \cite{FL01}. Analogous counting parameters for the 
case $M>1$ appear, within the context of wall polyominoes, in \cite{D99}. 
This study has been extended in \cite{NT03,NT03b} and in \cite{R05}. 
Whereas the focus in \cite{NT03,NT03b} is on particular examples 
of directed walks such as Dyck paths, \cite{D99} (resp.~\cite{R05}) 
also discuss the structure of underlying classes of functional equations 
in the case $M=1$ (resp.~$M>1$). Explicit expressions for the generating 
function $G(\boldsymbol u)$ have been obtained for $M=0$ and $M=1$, see the 
above references. A functional equation for staircase polygons,
counted by the above parameters and by width and height in addition 
(with formal variables $x$ and $y$), can be derived by the same method.
It is obtained from \eqref{form:fe} by multiplying the rhs with $xy$
and replacing the factor of two in the denominator by $(x+y)$.

\medskip

Instead of considering diagonal length moments, one can also
analyse {\it $k$-th column height moments} of a staircase polygon 
$p$, defined by $m_k(p)=\sum_{c\in C(p)} l(c)^k$, where the sum is 
over all (vertical) columns $c$ of $p$, with $C(p)$ denoting the 
set of (vertical) columns of $p$, and with $l(c)$ denoting the number of 
squares in column $c$. Obviously, $m_0(p)$ is the width, 
$m_1(p)$ the area of $p$. Counting polygons by column height 
moments, width and height $h(p)$ leads one to consider weights
\begin{equation*}
\widetilde w_p(\boldsymbol u,y)=y^{h(p)}\cdot u_0^{m_0(p)}\cdot\ldots\cdot u_M^{m_M(p)},
\end{equation*}
$y$ being a formal variable. Let $H(\boldsymbol u, y)=\sum_{p\in\cal P} 
\widetilde w_p(\boldsymbol u, y)$ denote the corresponding 
generating function. For a staircase polygon, consider the construction 
of moving the upper walk of the polygon one unit down, and removing its 
first edge and the last edge of the lower walk. Similarly to 
the case discussed above, this construction can be used to derive a 
functional equation for $H(\boldsymbol u, y)$. It is
\begin{equation}\label{form:anis}
H(\boldsymbol u,y)=y\frac{H(\boldsymbol v(\boldsymbol u),y)+u_0\cdot
\ldots\cdot u_M}{1-(H(\boldsymbol v(\boldsymbol u),y)+u_0\cdot\ldots
\cdot u_M)},
\end{equation}
where the functions $v_k({\boldsymbol u})$ are again given by 
\eqref{form:vk}.

Consider now the problem of counting staircase polygons by (total) 
half-perimeter and by column height moments. The generating function 
of the model is given by $H(\boldsymbol u, u_0)$. Since the sum of column 
heights of a staircase polygon is its area, for $M=1$ we have 
$H(u_0, u_1, u_0)=G(u_0,u_1)$ with functional
equation \eqref{form:fe}. For $M>1$, the diagonal moment approach for 
deriving a functional equation cannot be adapted to the present case,
since in the inflation step, which includes the addition of a vertical layer, 
the height of a column may be increased by more than one unit. It 
is not possible to keep track of this varying height increase without 
introducing new auxiliary variables in the generating function. But then, 
the method of extracting limit distributions of section 
\ref{limdist} cannot be applied in 
its present form. Since we prefer a model where limit distributions 
are considered with respect to large perimeter (and not to large width), 
we chose to discuss the diagonal length moment model. However, this choice is 
no restriction, since limit distributions of the first two models 
are of the same type, see section \ref{limdist}. Also, an exact 
analysis of the first few moments of the third model suggests the 
same type of limit distribution, see section \ref{uni} below.

\section{Perimeter and area generating function}

In this section, we will review the case $M=1$, which corresponds to
counting polygons by half-perimeter and area. Whereas most results discussed
below have appeared at various places in the literature 
\cite{PB95,D99,FL01,RGJ01,R02}, we recollect them here informally 
in order to prepare for the following section, where the case of general 
$M$ is discussed.

Consider the functional equation \eqref{form:fe} for $M=1$ and 
set $u_0=t$ and $u_1=q$. The generating function $G(t,q)$ of staircase 
polygons, counted by half-perimeter and area, satisfies the functional 
equation
\begin{equation}\label{form:pagf}
G(t,q)=\frac{t^2q}{1-2tq-G(tq,q)}.
\end{equation}
The function $G(t):=G(t,1)$ is the generating function for staircase
polygons counted by half-perimeter. If $q=1$, the functional 
equation \eqref{form:pagf} reduces to a 
quadratic equation, whose relevant solution is given by
\begin{equation*}
\begin{split}
G(t)&=\frac{1-2t-\sqrt{1-4t}}{2}
= \sum_{n=1}^\infty \frac{1}{n+1}\binom{2n}{n}t^{n+1}
= t^2 + 2t^3 + 5t^4 + 14t^5+ {\cal O}(t^6),
\end{split}
\end{equation*}
where $C_n=\binom{2n}{n}/(n+1)$ is the $n$-th Catalan number.
Thus, the number $p_n$ of staircase polygons of half-perimeter $n$
is given by
\begin{equation*}
p_n=C_{n-1}=\frac{1}{n}\binom{2n-2}{n-1} = \frac{1}{\sqrt{\pi}}
\frac{4^{n-1}}{n^{3/2}}\left( 1+{\cal O}(n^{-1})\right).
\end{equation*}
The above asymptotic form for $p_n$ can alternatively be obtained by
singularity analysis of the generating function $G(t)$. Such an 
approach will also be used in the following section. The 
function $t\mapsto G(t)$ is analytic for $|t|\le t_c=1/4$ 
except at $t=t_c$. The above expression yields an expansion 
about $t=t_c$ of the form
\begin{equation}\label{form:pgfex}
G(t) = G(t_c)+\sum_{l=0}^\infty f_{0,l} (t_c-t)^{(l+1)/2},
\end{equation}
where $G(t_c)=1/4$, $f_{0,0}=-1$, $f_{0,1}=1$, and $f_{0,l}=0$ for $l>1$.
In order to obtain information about the coefficients of $G(t)$,
we remind that functions $g(x)=(x_c-x)^{-\gamma}$ with rational 
exponent $\gamma\notin \{-1,-2,\ldots\}$ satisfy
\begin{equation}\label{form:excoeff}
[x^n]g(x)=\frac{1}{x_c^\gamma\Gamma(\gamma)}x_c^{-n}n^{\gamma-1}
\left(1+{\cal O}(n^{-1})\right),
\end{equation}
where $[x^n]f(x)$ denotes the coefficient of order $n$ in the 
power series $f(x)$, and $\Gamma(z)$ denotes the Gamma function 
\cite{FO90,SF05}. Applying this result to the (finite) expansion 
\eqref{form:pgfex} yields the above asymptotic form for $p_n$. 

We are interested in a probabilistic description of staircase 
polygons. Let ${\cal P}_{n}\subset \cal P$ denote the subset 
of staircase polygons of half-perimeter $n$. Consider a 
uniform model, where each polygon $p\in {\cal P}_{n}$ occurs 
with equal probability. We ask for the mean area of a polygon 
of half-perimeter $n$. To this end, introduce 
a discrete random variable $\widetilde X_n$ of polygon area by
\begin{equation*}
\mathbb P(\widetilde X_n=m)=
\frac{p_{n,m}}{\sum_m p_{n,m}},
\end{equation*}
where $p_{n,m}$ denotes the number of polygons of half-perimeter 
$n$ and area $m$. Note that the numbers $p_{n,m}$ appear in the
expansion $G(t,q)=\sum_{n,m}p_{n,m}t^nq^m$ of the perimeter and
area generating function $G(t,q)$. The mean area of a staircase 
polygon of half-perimeter $n$ is then given by
\begin{equation*} 
\mathbb E[\widetilde X_n]=\frac{\sum_m m p_{n,m}}{\sum_m p_{n,m}}
=\frac{[t^n]\left.q\frac{\partial}{\partial q}G(t,q)\right|_{q=1}}
{[t^n]G(t,1)}.
\end{equation*}
The function appearing in the numerator can be obtained from the functional
equation \eqref{form:pagf} by differentiating w.r.t.~$q$ and setting $q=1$.
This yields $\left.\frac{\partial}{\partial q}G(t,q)\right|_{q=1}=t^2/(1-4t)$,
and we get
\begin{equation*}
\mathbb E[\widetilde X_n] = \frac{4^{n-2}}{C_{n-1}}=\frac{\sqrt{\pi}}{4}n^{3/2}
\left(1+{\cal O}(n^{-1}) \right).
\end{equation*}
Thus, the mean area of a staircase polygon scales with its perimeter as 
$n^{3/2}$. For a single staircase polygon, the mean length of its diagonals 
is given by the quotient of area and half-perimeter $n$ minus one. Thus, the 
mean length of a diagonal of a random staircase polygon scales with its 
perimeter as $n^{1/2}$. The mean width of a staircase polygon may be analysed 
similarly using the width and height generating function, which is algebraic of
degree two, see \eqref{form:anis}. It can be shown that the mean width of a 
staircase polygon scales with its perimeter as $n$. Also, it can be shown that
the mean column height of a staircase polygon scales with its perimeter (width)
as $n^{1/2}$.

Higher moments of the area random variable $\widetilde X_n$ can be computed
similarly to the discussion above. For the $k$-th factorial moment of 
$\widetilde X_n$, where $k\in\mathbb N$, we have
\begin{equation}\label{form:pahmom}
\mathbb E[(\widetilde X_n)_k]=\frac{\sum_m (m)_kp_{n,m}}{\sum_m p_{n,m}}
=\frac{[t^n]\left.\frac{\partial^k}{\partial q^k}G(t,q)\right|_{q=1}}
{[t^n]G(t,1)},
\end{equation}
where $(a)_k=a(a-1)\cdot\ldots\cdot (a-k+1)$. Derivatives of $G(t,q)$
can be obtained recursively from the functional equation \eqref{form:pagf} 
by repeated differentiation w.r.t.~$q$. The result (see Lemma 
\ref{theo:Pui} in the next section) is given a follows. The derivative in 
\eqref{form:pahmom} is algebraic, with leading singular behaviour
\begin{equation}\label{form:agffk}
\frac{1}{k!}\left.\frac{\partial^k}{\partial q^k}G(t,q)\right|_{q=1}
=\frac{f_k}{(t_c-t)^{\gamma_k}}\left( 1+{\cal O}((t_c-t)^{1/2})\right),
\end{equation}
where $\gamma_k=3k/2-1/2$, and $f_k$ (strictly) positive. A recursion for 
the coefficient $f_k$ is given by
\begin{equation}\label{form:M1fk}
\frac{1}{16}\gamma_{k-1}f_{k-1}+\sum_{0\le l\le k} f_l f_{k-l}=0,
\end{equation}
where $k\in\mathbb N$ and $f_0=-1$. Inserting \eqref{form:agffk} into 
\eqref{form:pahmom} and extracting coefficients, using 
\eqref{form:excoeff} and standard transfer theorems \cite{FO90}, 
gives
\begin{displaymath}
\mathbb E[(\widetilde X_n)_k] =\frac{k!}{f_0t_c^{\gamma_k-\gamma_0}}
\frac{\Gamma(\gamma_0)}{\Gamma(\gamma_k)} f_k n^{3k/2}\left(1+{\cal O}(n^{-1/2})
\right).
\end{displaymath}
The (factorial) moments of $\widetilde X_n$ diverge for large perimeter. In order
to obtain a limit distribution, we introduce a normalised random variable $X_n$ 
of area by setting $X_n=\widetilde X_n/n^{3/2}$. This normalisation is natural 
in view of the scaling of the mean area with the perimeter, as described above.
Alternatively, the area is the sum over diagonal lengths (column heights) and thus
scales like $n\cdot n^{1/2}$, as argued above.  The moments of $X_n$ are then
asymptotically given by
\begin{displaymath}
\mathbb E[X_n^k]=\mathbb E[(X_n)_k]\left(1+{\cal O}(n^{-1/2})\right)
 =\frac{k!}{f_0t_c^{\gamma_k-\gamma_0}}
\frac{\Gamma(\gamma_0)}{\Gamma(\gamma_k)} f_k \left(1+{\cal O}(n^{-1/2})
\right),
\end{displaymath}
where we expressed the $k$-th moments in terms of the $r$-th factorial
moments, $r=0,1,\ldots,k$. The recursion \eqref{form:M1fk} can be used 
to show that the values $m_k=\lim_{n\to\infty} \mathbb E[X_n^k]$ satisfy 
the Carleman condition $\sum_k (m_k)^{-1/k}=+\infty$, implying the 
existence and uniqueness of a limit distribution with moments $m_k$ \cite{B86}. 
This distribution is known as the Airy distribution, see \cite{FL01} for 
a review. For models of simply generated trees, corresponding convergence 
results appear in \cite{A92,DM05}.

We finally describe a convenient technique to compute the numbers $f_k$, 
which will be central in the following sections. This approach, called the
method of dominant balance \cite{PB95}, was initially
followed when analysing the scaling function of staircase polygons
\cite{RGJ01}. Whereas it has been applied under a certain analyticity 
assumption on the corresponding model, it can be given a rigorous meaning 
at a formal level, see the remark after Lemma \ref{theo:Pui} in the following 
section. It has been proved \cite[Thm.~5.3]{P95} that there 
exists a uniform asymptotic expansion of the perimeter and area generating
function $G(t,q)$ about $t=t_c$, which is, to leading order, given by
\begin{equation}\label{form:scalans}
G(t,q)\sim G(t_c)+ (t_c-t)^{1/2} F_0\left( \frac{1-q}{(t_c-t)^{3/2}}\right)
\qquad (t\to t_c^-),
\end{equation}
uniformly in $q\le1$ near unity. The function $F_0(\epsilon)$ is called 
the scaling function and admits an asymptotic expansion about $\epsilon=0$,
the exponents $1/2$ and $3/2$ are called the critical exponents of the model, 
see also \cite{PB95,P95,J00}. The scaling function is explicitly given by
\begin{equation*}
F_0(\epsilon)=\epsilon^{1/3}\frac{2^{-4/3}\mbox{Ai}'(2^{8/3}\epsilon^{-2/3})}
{\mbox{Ai}(2^{8/3}\epsilon^{-2/3})}
=-1-\frac{1}{64}\epsilon+\frac{5}{8192}\epsilon^2-\frac{15}{262144}\epsilon^3+
{\cal O}(\epsilon^4),
\end{equation*}
where $\mbox{Ai}(z)=\frac{1}{\pi}\int_0^\infty\cos(t^3/3+tz){\rm d}t$ is 
the Airy function. The coefficients $f_k$ of \eqref{form:agffk}
appear in the scaling function via $F_0(\epsilon)=\sum_{k\ge0} (-1)^k 
f_k \epsilon^k$. This is seen by asymptotically evaluating the lhs 
of \eqref{form:agffk}, using the scaling form \eqref{form:scalans}, and
by comparing to the rhs of \eqref{form:agffk}. The scaling function can be
extracted from the functional equation \eqref{form:pagf} as follows. Insert 
the scaling form \eqref{form:scalans} for the generating function $G(t,q)$
in the functional equation \eqref{form:pagf} and introduce new variables $s$ and 
$\epsilon$ by setting $x=x_c-s^2$ and $q=1-\epsilon s^3$. Then, the
functional equation yields a differential equation for $F_0(\epsilon)$, 
when expanded up to order $s^2$, see also \cite{R02} for a detailed account. 
The differential equation for $F_0(\epsilon)$ is of Riccati type and given by
\begin{equation*}
\frac{1}{16}\epsilon\left( \frac{1}{2}F_0(\epsilon)-\frac{3}{2}\epsilon 
F_0'(\epsilon)\right) +F_0(\epsilon)^2=1.
\end{equation*}
This immediately translates into the above recursion \eqref{form:M1fk} for 
the coefficients $f_k$ of the scaling function. 

\section{Limit distributions and scaling functions}\label{limdist}

We now consider the case of arbitrary $M$. Whereas this leads to cumbersome
expressions, we are following the same ideas as in the previous
section. Recall that the set of staircase polygons of half-perimeter $n_0$ 
is denoted by ${\cal P}_{n_0}$. We  consider a uniform model where each polygon 
$p\in {\cal P}_{n_0}$ occurs with equal probability.
Then, to each counting parameter a corresponding random variable is attached. This 
leads us to consider discrete random variables $\widetilde X_{k,n_0}$ for the 
diagonal length moments $n_k(p)$, for $k=1,\ldots,M$. The generating function 
$G(\boldsymbol u)$ is a formal power series in $\boldsymbol u$, i.e.,  
$G(\boldsymbol u)\in\mathbb C[[\boldsymbol u]]$, where $R[[x_1,\ldots,x_N]]$ 
denotes the ring of formal $R$-power series in the variables $x_1,\ldots, 
x_N$. In fact, $G(\boldsymbol u)\in \mathbb Z[u_1,\ldots,u_M][[u_0]]$, 
where $R[x_1,\ldots,x_N]$ denotes the ring of $R$-polynomials in the variables 
$x_1,\ldots, x_N$.
The function $G(\boldsymbol u_0)$, where $\boldsymbol u_0=(u_0,1,\ldots,1)$,
as considered in the previous section, is the (half-) perimeter generating
function of the model. Write the generating function $G(\boldsymbol u)$ as
\begin{equation*}
G(\boldsymbol u)=\sum_{n_0,n_1,\ldots, n_M} p_{n_0,n_1,\ldots,n_M} u_0^{n_0}u_1^{n_1}
\cdot\ldots\cdot u_M^{n_M}.
\end{equation*}
The coefficient $p_{n_0,n_1,\ldots,n_M}$ is the number of staircase polygons of
half-perimeter $n_0$ where, for $k=1,\ldots,M$, the $k$-th diagonal length 
moment has the value $n_k$. We have $0<\sum_{n_1,\ldots, n_M}p_{n_0,n_1,
\ldots,n_M}=[u_0^{n_0}]G(\boldsymbol u_0)=C_{n_0-1}<\infty$, where
$C_n$ is the $n$-th Catalan number. In a uniform model, where each polygon 
$p\in {\cal P}_{n_0}$ has the same probability $1/C_{n_0-1}$, the
discrete random variables $\widetilde X_{k,n_0}$ corresponding to the 
diagonal moments $n_k(p)$ have the probability distribution
\begin{equation*}
\mathbb P(\widetilde X_{1,n_0}=n_1, \ldots, \widetilde X_{M,n_0}=n_M)=
\frac{p_{n_0,n_1,\ldots,n_M}}{\sum_{n_1,\ldots, n_M}p_{n_0,n_1,\ldots,n_M}}.
\end{equation*}
We will analyse the joint distribution of the random variables 
$\widetilde X_{k,n_0}$, where $k=1,\ldots, M$, in the limit of large perimeter.
This will be achieved by studying asymptotic properties of mixed moments 
of the random variables $\widetilde X_{k,n_0}$, given by
\begin{equation*}
\mathbb E[\widetilde X_{1,n_0}^{k_1}\cdot\ldots\cdot
\widetilde X_{M,n_0}^{k_M}]=\frac{\sum_{n_1,\ldots, n_M} n_1^{k_1}
\cdot\ldots\cdot n_M^{k_M}p_{n_0,n_1,\ldots,n_M}}{\sum_{n_1,\ldots, n_M}
p_{n_0,n_1,\ldots,n_M}}.
\end{equation*}
We employ a generating function method. Note that we have
\begin{equation*}
\mathbb E[\widetilde X_{1,n_0}^{k_1}\cdot\ldots\cdot
\widetilde X_{M,n_0}^{k_M}]=
\frac{[u_0^{n_0}]\left.\left(u_1\frac{\partial}{\partial u_1}\right)^{k_1}
\cdot\ldots\cdot \left(u_M\frac{\partial}{\partial u_M}\right)^{k_M} 
G(\boldsymbol u)\right|_{\boldsymbol u =\boldsymbol u_0}}{[u_0^{n_0}]
G(\boldsymbol u_0)}.
\end{equation*}
We can thus study asymptotic properties of the moments by studying
asymptotic properties of the coefficients of the generating functions 
appearing on the rhs of the above equation. The latter problem can be
treated by singularity analysis of generating functions, as indicated 
in the previous section. It will prove convenient to consider factorial 
moments, given by
\begin{equation*}
\mathbb E[(\widetilde X_{1,n_0})_{k_1}\cdot\ldots\cdot
(\widetilde X_{M,n_0})_{k_M}]=
\frac{[u_0^{n_0}]\left.\frac{\partial^{k_1}}{\partial u_1^{k_1}}
\cdot\ldots\cdot \frac{\partial^{k_M}}{\partial u_M^{k_M}} 
G(\boldsymbol u)\right|_{\boldsymbol u =\boldsymbol u_0}}{[u_0^{n_0}]
G(\boldsymbol u_0)}.
\end{equation*}
Moments can be expressed linearly in factorial moments. It will turn out 
below that they are, for $k_1,\ldots, k_M$ fixed, asymptotically equal. 
Let us introduce {\it factorial moment generating functions} 
$g_{\boldsymbol k}(u_0)$, defined by
\begin{equation*}
g_{\boldsymbol k}(u_0):= \left.\frac{1}{\boldsymbol k!}
\frac{\partial^{k_1}}{\partial u_1^{k_1}}\cdots \frac{\partial^{k_M}}
{\partial u_M^{k_M}} G(\boldsymbol u)
\right|_{\boldsymbol u =\boldsymbol u_0},
\end{equation*}
where $\boldsymbol  k=( k_1,\ldots, k_M)\in\mathbb N_0^M$. We use the 
multi-index notation. In particular, $\boldsymbol k!=k_1!\cdot\ldots
\cdot k_M!$, $|\boldsymbol k|=k_1+\ldots+k_M$, and $\boldsymbol k\le
\boldsymbol l$ if $k_i\le l_i$ for $i=1,\ldots,M$. We will also 
use unit vectors $\boldsymbol e_k$, $k=1,\ldots, M$, with 
coordinates $(e_k)_i =\delta_{i,k}$ for $i=1,\ldots, M$.
We clearly have $g_{\boldsymbol k}(u_0)\in\mathbb C[[u_0]]$ for all 
$\boldsymbol k$. The function $g_{\boldsymbol  0}(u_0)=G(\boldsymbol 
u_0)=G(u_c)+\sum_{l=0}^\infty f_{\boldsymbol 0,l}(u_c-u_0)^{(l+1)/2}$ has been
studied in the previous section \eqref{form:pgfex}, where we found 
$u_c=1/4$, $G(u_c)=1/4$, $f_{\boldsymbol 0,0}=-1$, $f_{\boldsymbol 0,1}=1$, and 
$f_{\boldsymbol 0,l}=0$ for $l>1$. The functional equation \eqref{form:fe} 
can be used to show that the properties of $g_{\boldsymbol  0}(u_0)$ carry 
over to those of $g_{\boldsymbol  k}(u_0)$ for $\boldsymbol 
k\ne\boldsymbol0$. We have the following lemma.

\begin{lemma}\label{theo:Pui}
For $\boldsymbol  k\ne\boldsymbol0$, all factorial moment generating functions 
$g_{\boldsymbol  k}(u_0)$ are algebraic. They are analytic for $|u_0|\le u_c=1/4$, 
except at $u_0=u_c$, with Puiseux expansions of the form

\begin{equation}\label{form:gk}
g_{\boldsymbol k}(u_0) = \sum_{l=0}^\infty f_{\boldsymbol  k,l}(u_c-u_0)^{l/2-\gamma_{\boldsymbol k}},
\end{equation}
where $\gamma_{\boldsymbol k}=-1/2+\sum_{i=1}^M(1+i/2) k_i$. The leading coefficients
$f_{\boldsymbol k,0}=f_{\boldsymbol k}$ are given by 
$f_{\boldsymbol k}=c_{\boldsymbol k}f_{\boldsymbol 0}^{1-|\boldsymbol k|}f_{\boldsymbol e_1}^{k_1}
\cdot\ldots\cdot f_{\boldsymbol e_M}^{k_M}$, where the numbers $c_{\boldsymbol k}$ are, for $\boldsymbol 
k\ne \boldsymbol 0$, determined by the recursion
\begin{equation}\label{form:fk}
c_{\boldsymbol k} = -2\gamma_{{\boldsymbol k}-{\boldsymbol e}_1}
c_{{\boldsymbol k}-{\boldsymbol e}_1}+\sum_{i=1}^{M-1}( k_i+1)
c_{{\boldsymbol k}-{\boldsymbol e}_{i+1}+{\boldsymbol e}_i}-\frac{1}{2}
\sum_{\substack{{\boldsymbol l}\ne {\bf 0},{\boldsymbol l}
\ne {\boldsymbol k}\\ {\bf 0}\le{\boldsymbol l}\le{\boldsymbol k}}}
c_{\boldsymbol l}c_{{\boldsymbol k}-{\boldsymbol l}},
\end{equation}
with boundary conditions $c_{\boldsymbol 0}=1$ and $c_{\boldsymbol k}=0$ if 
$ k_j<0$ for some $1\le j\le M$. The coefficients $f_{\boldsymbol k}$ are strictly 
positive for ${\boldsymbol k}\ne {\bf 0}$. Explicitly, we have 
$f_{\boldsymbol 0}=-1$ and $f_{\boldsymbol e_k}=k! 2^{-3(k+1)}$ for $k=1,\ldots,M$. 
\end{lemma}

\begin{proof}
The above lemma is a special case of \cite[Prop.~5.3,~5.6]{R05}. We outline the
idea of proof. 

Successively differentiating the functional equation \eqref{form:fe} yields
$g_{\boldsymbol k}(u_0)$ as rational function of $g_{\boldsymbol 0}(u_0)$
and its derivatives. Due to the closure properties of
algebraic functions, these functions are algebraic again. The functions
$g_{\boldsymbol k}(u_0)$ then inherit analytic properties of 
$g_{\boldsymbol 0}(u_0)$. In particular, $g_{\boldsymbol k}(u_0)$
is analytic for $|u_0|\le u_c$, except at $u_0=u_c$, with a Puiseux expansion
of the form \eqref{form:gk} for some exponent $\widetilde \gamma_{\boldsymbol k}$. 
The value of the leading exponent $\gamma_{\boldsymbol k}$ and the recursion 
for the coefficients $f_{\boldsymbol k}$ (hence for the coefficients 
$c_{\boldsymbol k}$) is proved by induction, via asymptotically analysing 
the $\boldsymbol k$-th derivative of the functional equation.
\end{proof}

\noindent {\bf Remarks.}\\
{\it i)} For the model \eqref{form:anis} of staircase polygons, counted 
by width and (vertical) column height moments, Lemma \ref{theo:Pui} applies 
as well, however with different numbers $u_c=(1-\sqrt{y})^2$ for $0<y<1$, 
$f_{\boldsymbol 0}=-y^{-1/4}$ and $f_{\boldsymbol e_k}=k!2^{-k}y^{-k/4}$ 
for $k=1,\ldots, M$. It has been shown that Lemma \ref{theo:Pui} applies 
in a much more general context \cite[Prop.~5.6]{R05}. In particular, the 
amplitude ratios $f_{\boldsymbol k}f_{\boldsymbol 0}^{|\boldsymbol k|-1}
f_{\boldsymbol e_1}^{-k_1}\cdot\ldots\cdot f_{\boldsymbol e_M}^{-k_M}=
c_{\boldsymbol k}$ are model independent in that situation, i.e., independent
of $u_c$, $f_{\boldsymbol 0}$ and $f_{\boldsymbol e_k}$.\\
{\it ii)} The recursion \eqref{form:fk} for the coefficients 
$f_{\boldsymbol k}$ can be obtained mechanically, if a tight bound on the 
exponents $\gamma_{\boldsymbol k}$ is known. (Typically, analysis of the 
first few functions $g_{\boldsymbol k}(u_0)$ suggests the form of 
$\gamma_{\boldsymbol k}$. Then, one can prove that $\gamma_{\boldsymbol 
k}$ is an exponent bound by induction, using the functional equation.
Such a proof is generally easier than a full asymptotic analysis of the
functional equation, see \cite[Prop.~5.5]{R05}.)
If we replace the functions $g_{\boldsymbol k}(u_0)$ in the expansion of 
$G(\boldsymbol u)$ about $u_1=\ldots=u_M=1$ by their Puiseux expansions 
\eqref{form:gk}, we then have
\begin{equation}\label{form:sf}
\begin{split}
G(\boldsymbol u)&=\sum_{\boldsymbol k} (-1)^{|\boldsymbol k|} g_{\boldsymbol k}(u_0) 
(1-u_1)^{k_1}\cdot\ldots\cdot (1-u_M)^{k_M}\\
&= G(u_c)+\sqrt{u_c-u_0} F\left(\frac{1-u_1}{(u_c-u_0)^{(1+2)/2}}, \ldots, 
\frac{1-u_M}{(u_c-u_0)^{(M+2)/2}}, \sqrt{u_c-u_0}\right),
\end{split}
\end{equation}
where, for $\boldsymbol \epsilon=(\epsilon_1,\ldots,\epsilon_M)$, the function  
$F(\boldsymbol \epsilon,s)\in\mathbb C[[\boldsymbol \epsilon,s]]$ is a formal power series, 
and $F(\boldsymbol \epsilon,s)=\sum_l F_l(\boldsymbol \epsilon)s^l$,
where $F_l(\boldsymbol \epsilon)=\sum_{\boldsymbol k}(-1)^{|\boldsymbol k|}f_{\boldsymbol k, l}
\boldsymbol \epsilon^{\boldsymbol k}$. In statistical mechanics, the function 
$F_0(\boldsymbol\epsilon)$ is called the scaling function, and the functions 
$F_l(\boldsymbol \epsilon)$ are called the correction-to-scaling functions, if the equation
\eqref{form:sf} is valid as an asymptotic expansion in $u_c-u_0$, uniformly in 
$1-q_1,\ldots,1-q_M$, see \cite{J00}. For a probabilistic interpretation, 
see the remark after Theorem \ref{theo:exc}.

The functional equation \eqref{form:fe} induces partial differential equations for the
functions $F_l(\boldsymbol \epsilon)$. They are obtained by substituting \eqref{form:sf}
in \eqref{form:fe}, and by introducing variables $s,\epsilon_1,\ldots, \epsilon_M$ via 
$u_0=u_c-s^2$ and $u_k=1-\epsilon_k s^{k+2}$ for $k=1,\ldots, M$. Then, a partial 
differential equation for $F_l(\boldsymbol \epsilon)$ appears in the expansion of 
the functional equation in $s$ at order $s^{l+2}$. Computing these equations is 
called the method of dominant balance \cite{PB95}, see \cite{R02} for the case $M=1$,
and \cite{R05} for the case of general $M$. For example, we get for 
$F_0(\boldsymbol \epsilon)$ the equation
\begin{equation}\label{form:pde}
\frac{1}{16}\epsilon_1\left(\frac{1}{2}F_0(\boldsymbol \epsilon) -
\sum_{i=1}^M \left(1+\frac{i}{2}\right)\epsilon_i 
\frac{\partial F_0}{\partial \epsilon_i}(\boldsymbol\epsilon)\right)
+\sum_{i=1}^{M-1}\frac{(i+1)}{4}\epsilon_{i+1}\frac{\partial F_0}
{\partial \epsilon_i}(\boldsymbol\epsilon)+F_0(\boldsymbol\epsilon)^2=1.
\end{equation}
This yields, for $\boldsymbol k\ne\boldsymbol 0$, a recursion for the coefficients 
$f_{\boldsymbol k}$ of $F_0(\boldsymbol\epsilon)$ of the form
\begin{displaymath}
\frac{1}{16}\gamma_{{\boldsymbol k}-{\boldsymbol e}_1}
f_{{\boldsymbol k}-{\boldsymbol e}_1}+\sum_{i=1}^{M-1}\frac{i+1}{4}( k_i+1)
f_{{\boldsymbol k}-{\boldsymbol e}_{i+1}+{\boldsymbol e}_i}+
\sum_{\bf 0\le{\boldsymbol l}\le{\boldsymbol k}}
f_{\boldsymbol l}f_{{\boldsymbol k}-{\boldsymbol l}}=0,
\end{displaymath}
which translates into the above recursion \eqref{form:fk} for the coefficients 
$c_{\boldsymbol k}$.

\medskip

We are now prepared for a discussion of moment asymptotics. Recall that
the random variables $\widetilde X_{k,n_0}$ are related to the diagonal 
height moments $n_k(p)$, which are defined in \eqref{form:dm}. Since the 
number of diagonals of a staircase polygon equals its half-perimeter
minus one, and the average diagonal length of a staircase polygon scales 
like $n_0^{1/2}$, we expect that the mean $k$-th diagonal length 
moment scales like $n_0^{1+k/2}$. We thus introduce normalised random 
variables $X_{k,n_0}$ by 
\begin{equation}\label{form:nrv}
(X_{1,n_0},X_{2,n_0},\ldots, X_{M,n_0})= \left(\frac{\widetilde X_{1,n_0}}{n_0^{3/2}},
\frac{\widetilde X_{2,n_0}}{n_0^{4/2}},\ldots, 
\frac{\widetilde X_{M,n_0}}{n_0^{(M+2)/2}}\right).
\end{equation}
Then, the mixed moments of the random variables $X_{k,n_0}$ converge to a 
finite, nonzero limit as $n_0\to \infty$. We have the following result.

\begin{theorem}\label{theo:exc}
The random variable $(X_{1,n_0},X_{2,n_0},\ldots, X_{M,n_0})$
of eqn.~\eqref{form:nrv} converges in distribution,
\begin{displaymath}
(X_{1,n_0},X_{2,n_0},\ldots, X_{M,n_0}) \stackrel{d}{\to} (\alpha_1X_1,\ldots, \alpha_M X_M),
\end{displaymath}
where $X_k=\int e^k(t)\,{\rm d}t$ denotes the integral of the $k$-th power of the standard Brownian 
excursion of duration $1$, and the numbers $\alpha_k$ are given by 
$\alpha_k=-\frac{f_{\boldsymbol e_k}}{f_{\boldsymbol 0}}\frac{2^{3-3k/2}}{k!}$.
We also have moment convergence with
\begin{displaymath}\label{form:exc1}
\mathbb E[X_{1,n_0}^{k_1}\cdot\ldots\cdot X_{M,n_0}^{k_M}]=
\frac{\boldsymbol k!}
{f_{\bf 0}u_c^{\gamma_{\boldsymbol k}-\gamma_{\bf 0}}}
\frac{\Gamma(\gamma_{\boldsymbol 0})}{\Gamma(\gamma_{\boldsymbol k})} 
f_{\boldsymbol k} \left( 1+{\cal O}(n_0^{-1/2})\right),
\end{displaymath}
where the numbers $u_c$, $\gamma_{\boldsymbol k}$ and 
$f_{\boldsymbol k}$ are those of Lemma \ref{theo:Pui}. The numbers 
$\alpha_k$ are explicitly given by $\alpha_k=2^{-9k/2}$.
\end{theorem}

\begin{proof}
Theorem \ref{theo:exc} is obtained by specialising \cite[Thm.~8.5]{R05} 
and \cite[Cor.~9.1]{R05} to the functional equation \eqref{form:fe}. We sketch the proof.

By standard transfer theorems \cite{FO90} for algebraic functions, 
the results of Lemma \ref{theo:Pui} translate to an asymptotic 
expression for the factorial moments. It is
\begin{displaymath}
\mathbb E[(\widetilde X_{1,n_0})_{k_1}\cdot\ldots\cdot(\widetilde 
X_{M,n_0})_{k_M}]=
\frac{{\boldsymbol k}!}{f_{\bf 0}
u_c^{\gamma_{\boldsymbol k}-\gamma_{\bf 0}}}
\frac{\Gamma(\gamma_{\bf 0})}{\Gamma(\gamma_{\boldsymbol k})}f_{\boldsymbol k}
n_0^{\gamma_{\boldsymbol k}-\gamma_{\bf 0}}+{\cal O}(n_0^{\gamma_{\boldsymbol k}-
\gamma_{\bf 0}-1/2}),
\end{displaymath}
The explicit error term implies that factorial 
moments are asymptotically equal to the (ordinary) moments $\mathbb 
E[\widetilde X_{1,n_0}^{k_1}\cdot\ldots\cdot\widetilde X_{M,n_0}^{k_M}]$. 
In particular, the moments $\mathbb E[X_{1,n_0}^{k_1}\cdot\ldots\cdot 
X_{M,n_0}^{k_M}]$ converge to a finite, nonzero limit as $n_0\to
\infty$. We have
\begin{displaymath}
\mathbb E[X_{1,n_0}^{k_1}\cdot\ldots\cdot X_{M,n_0}^{k_M}]=
\frac{\mathbb E[\widetilde X_{1,n_0}^{k_1}\cdot\ldots\cdot
\widetilde X_{M,n_0}^{k_M}]}{n_0^{\gamma_{\boldsymbol k}-\gamma_{\bf 0}}}
= \frac{{\boldsymbol k}!}{f_{\bf 0}u_c^{\gamma_{\boldsymbol k}-\gamma_{\bf 0}}}
\frac{\Gamma(\gamma_{\bf 0})}{\Gamma(\gamma_{\boldsymbol k})}f_{\boldsymbol k}
\left( 1+{\cal O}(n_0^{-1/2})\right).
\end{displaymath}
Set $m_{\boldsymbol k}=\lim_{n_0\to\infty} \mathbb E[X_{1,n_0}^{k_1}
\cdot\ldots\cdot X_{M,n_0}^{k_M}]$. The recursion \eqref{form:fk} or
the partial differential equation \eqref{form:pde} induces an upper 
bound on the growth of the coefficients $f_{\boldsymbol k}$. This 
can be used to show that for $\boldsymbol t\in\mathbb R^M$ 
\begin{displaymath}
\lim_{|\boldsymbol k|\to\infty} 
\frac{m_{\boldsymbol k}\boldsymbol t^{\boldsymbol k}}{\boldsymbol k!}=0.
\end{displaymath}
This guarantees the existence and uniqueness of a limit distribution with
moments $m_{\boldsymbol k}$ via L\'evy's continuity theorem \cite{B96}. 
The connection to Brownian excursions follows by comparing the moments, 
see \cite[Thm.~2.4]{R05}.
\end{proof}

\noindent {\bf Remarks.} \\
{\it i)} For a discussion of the case $M=1$, see the previous section. 
The case $M=2$ has been analysed in \cite{NT03,NT03b}.
The above approach yields the existence and uniqueness of a (joint)
limit distribution for $M$ arbitrary, and provides recurrences for
their mixed moments. It also allows to analyse corrections to the
limiting behaviour. It applies in a more general setup of models
whose generating function satisfy a $q$-functional equation 
\cite{R05}. This includes simply generated trees as a special case 
\cite{A92,DM05}. For $M=1$ and $M=2$, moment recurrences for Dyck paths 
have also been obtained \cite{NT03b} via Louchard's formula \cite{L84} 
from stochastics. The function $F_0(\boldsymbol \epsilon)$ in \eqref{form:sf} 
is a certain Laplace transform of the moment generating function for the 
limiting probability distribution, see \cite[Thm.~7]{NT03}.\\
{\it ii)} For the model \eqref{form:anis} of staircase polygons, counted 
by width and vertical column height moments, Theorem \ref{theo:exc} applies 
as well, however with different numbers $u_c$, $f_{\boldsymbol 0}$ and 
$f_{\boldsymbol e_k}$, see the remark after Lemma \ref{theo:Pui}. The 
numbers $\alpha_k$ are, for $k=1,\ldots,M$, explicitly given by 
$\alpha_k=y^{(k-1)/4}2^{3-5k/2}$. The same remark also implies that the
ratios
\begin{equation}\label{form:erwratios}
\frac{\mathbb E[X_{1,n_0}^{k_1}\cdot\ldots\cdot X_{M,n_0}^{k_M}]}
{\mathbb E[X_{1,n_0}]^{k_1}\cdot\ldots\cdot\mathbb E[{X_M,n_0}]^{k_M}}=
\frac{\boldsymbol k! c_{\boldsymbol k} \Gamma(\gamma_{\boldsymbol 0})^{1-
|\boldsymbol k|} 
\Gamma(\gamma_{{\boldsymbol e}_1})^{k_1}\cdot\ldots\cdot
\Gamma(\gamma_{{\boldsymbol e}_M})^{k_M}}{ \Gamma(\gamma_{{\boldsymbol k}})}
\left( 1+{\cal O}\left( n_0^{-1/2}\right)\right).
\end{equation}
are asymptotically model independent, i.e., asymptotically independent of $u_c$, 
$f_{\boldsymbol 0}$, and $f_{{\boldsymbol e}_k}$, where $k=1,\ldots, M$. 
In particular, the corresponding numbers coincide for staircase polygons, 
counted by perimeter and diagonal length moments, and for staircase polygons 
counted by width and column height moments. For this reason, we will use 
the ratios \eqref{form:erwratios} to check for 
a particular type of limit distribution in the following section.

\section{Universality and self-avoiding polygons}\label{uni}

\begin{figure}[htb]
\begin{center}
\begin{minipage}[b]{0.95\textwidth}
\center{\epsfig{file=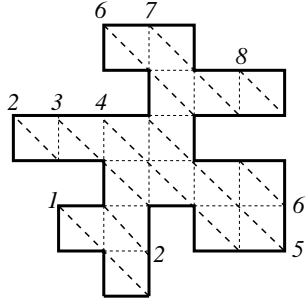,width=4cm}}
\end{minipage}
\end{center}
\caption{\label{fig:sapcol}
\small A self-avoiding polygon with eight diagonal layers, which
each consist of one or several connected segments. The second and sixth 
diagonal layer from below each consist of two diagonal segments. The other
diagonal layers each consist of a single segment.}
\end{figure}

For self-avoiding polygons, the notion of diagonal length moments 
as studied in the previous sections is not well-defined, since the 
intersection of a diagonal line with the (area of the) polygon
need not be connected. However, we can consider a self-avoiding polygon $p$ of
finite perimeter as being built from a set of diagonal layers $DL(p)$, where 
each layer consists of a finite number of connected segments $d$, see 
Figure \ref{fig:sapcol}. If we restrict to staircase polygons, each layer 
consists of a single segment, see Figure \ref{fig:diag}. We define 
parameters $n_k^{(a)}(p)$ and $n_k^{(b)}(p)$ by
\begin{equation*}
n_k^{(a)}(p)=\sum_{L\in DL(p)} \left(\sum_{d\in L} l(d)\right)^k,\qquad 
n_k^{(b)}(p)=\sum_{L\in DL(p)} \left(\sum_{d\in L} l(d)^k\right),
\end{equation*}
where $l(d)$ is the length of a segment $d$ of a diagonal layer $L$, 
see also Figure \ref{fig:sapcol}. 
Let us call these parameters {\it $k$-th diagonal layer moments}.
For staircase polygons $p$, we have $n_k(p)=n_k^{(a)}(p)=n_k^{(b)}(p)$. For
self-avoiding polygons $p$, we have $n_1^{(a)}(p)=n_1^{(b)}(p)$, and 
$n^{(a)}(p)$ is the area of the polygon.
The case $k=1$ has been studied in a number of different investigations 
\cite{RGJ01,RJG03}, supporting the assumption that the limit distribution
of staircase polygon area and self-avoiding polygon area are of the
same type. Apparently, the above parameters have not been studied previously 
for $k>1$. We estimated moments of the parameters $n_k^{(a)}(p)$ and $n_k^{(b)}(p)$ 
by a Monte-Carlo simulation of self-avoiding polygons, within a
uniform model where each polygon of fixed perimeter occurs with equal 
probability. The algorithm is described in \cite{MOS90}. Each Monte-Carlo 
step consists of an inversion or a certain reflection of a randomly chosen 
part of the self-avoiding polygon. Polygons which are no longer self-avoiding 
are rejected. Checking for self-avoidance is done in time proportional to the 
polygon length. Determination of the segment structure of diagonal layers is 
done ``on the fly'' in time proportional to the polygon length. For a given 
half-perimeter $n_0$, we took a sample of $10^6$ polygons, with at least 
$10\times n_0$ Monte-Carlo update moves between consecutive measurements. 
We used the random number generator {\tt ran2} described in \cite{PTVF92}.
Estimates for the $r$-th moment of the random variable corresponding to the
parameter $n_k^{(a)}$ are, for fixed half-perimeter $n_0$, given by
\begin{equation*}
m^{(a)}_{k,n_0}(r) := \frac{\sum_n n^r \widetilde p_{n_0,n}}{\sum_n 
\widetilde p_{n_0,n}},
\end{equation*}
where $\widetilde p_{n_0,n}$ denotes the number of sampled self-avoiding
polygons of half-perimeter $n_0$ and value $n$ of the parameter $n_k^{(a)}$.
An analogous expression is used for the $r$-th moment $m^{(b)}_{k,n_0}(r)$ 
of the random variable corresponding to the parameter $n_k^{(b)}$. If the 
parameters $n_k^{(a)}$, $n_k^{(b)}$, and $n_k$ have limit distributions of 
the same type, then the moment ratios $m^{(a)}_{k,n_0}(r)/m^{(a)}_{k,n_0}(1)^r$,
$m^{(b)}_{k,n_0}(r)/m^{(b)}_{k,n_0}(1)^r$ and 
$\mathbb E[\widetilde X^r_{k,n_0}]/\mathbb E[\widetilde X_{k,n_0}]^r$ are
asymptotically equal, see the remark after Theorem \ref{theo:exc}.

For comparison, we present data of the case $k=1$ for the second area moment.
Its asymptotic moment ratio is, for staircase polygons, given by
\begin{equation*}
\lim_{n_0\to\infty}\frac{\mathbb E[(\widetilde X_{1,n_0})^2]}
{\mathbb E[\widetilde X_{1,n_0}]^2} = \frac{10}{3\pi}\approx 1.06103
\end{equation*}
\begin{figure}[htb]
\begin{center}
\begin{minipage}[b]{0.95\textwidth}
\center{\epsfig{file=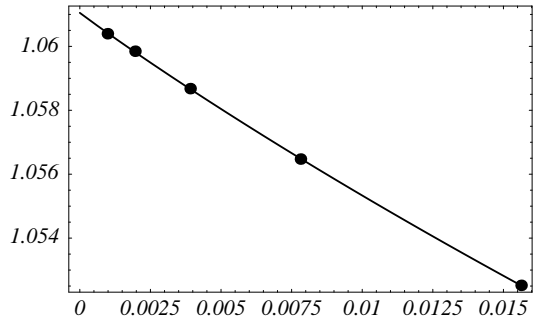,width=7cm}}
\end{minipage}
\end{center}
\caption{\label{fig:r1}
\small Plot of the moment ratio $m^{(a)}_{1,n_0}(2)/m^{(a)}_{1,n_0}(1)^2$
against $1/(2n_0)$. The line is a least square fit through the data points.}
\end{figure}
We sampled self-avoiding polygons for perimeter values $2n_0\in\{64, 128,
256512, 1024\}$ and extrapolated the asymptotic moment ratio by a least 
square fit, obtaining the value $1.06084$, see Figure \ref{fig:r1}. Both 
the value of the least square fit and the small data spread in the figure 
indicate that this value is consistent with the corresponding number for 
staircase polygons. 

\begin{figure}[htb]
\begin{center}
\begin{minipage}[b]{0.8\textwidth}
\epsfig{file=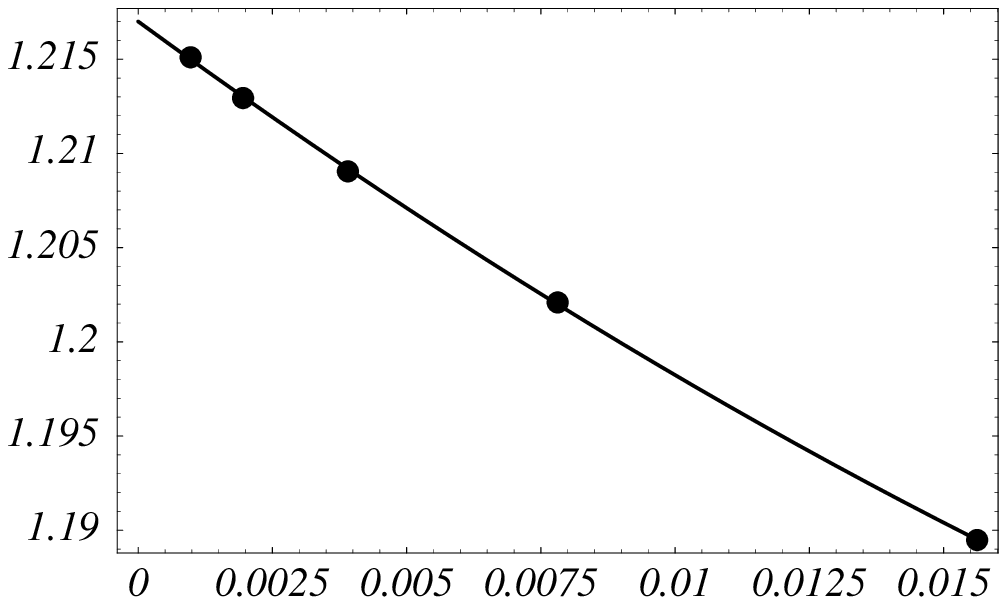,width=6cm}
\hfill
\epsfig{file=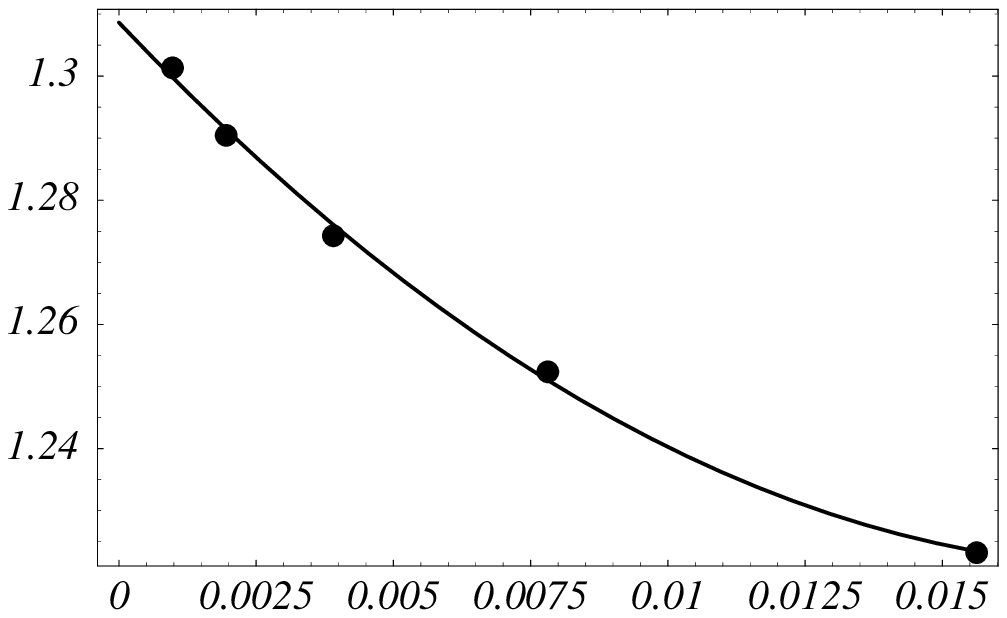,width=6cm}
\end{minipage}
\end{center}
\caption{\label{fig:r2}
\small  Plots of the moment ratios $m^{(a)}_{2,n_0}(2)/m^{(a)}_{2,n_0}(1)^2$
(left) and $m^{(b)}_{2,n_0}(2)/m^{(b)}_{2,n_0}(1)^2$ (right) against $1/(2n_0)$.
The lines are least square fits through the data points.}
\end{figure}

For $k=2$, we present data for the second diagonal layer moment in Figure 
\ref{fig:r2},
using the two different definitions $n_2^{(a)}$ and  $n_2^{(b)}$. 
We sampled self-avoiding polygons for perimeter values $2n_0\in\{64, 128,
256, 512, 1024\}$. Extrapolating the asymptotic moment ratios by a
least square fit yields the values $m^{(a)}_{2,n_0}(2)/m^{(a)}_{2,n_0}(1)^2
\to1.2162$ and  $m^{(b)}_{2,n_0}(2)/m^{(b)}_{2,n_0}(1)^2\to1.3088$.
For staircase polygons, the corresponding ratio is
\begin{equation*}
\lim_{n_0\to\infty}\frac{\mathbb E[(\widetilde X_{2,n_0})^2]}
{\mathbb E[\widetilde X_{2,n_0}]^2} = \frac{19}{15}\approx 1.26667
\end{equation*}
This indicates different limit distributions for both models, in 
contrast to the case $k=1$.

We also considered a corresponding generalisation of column height
moments, which we call vertical layer moments. In section \ref{sec:feq},
we analysed the model of staircase polygons, counted by width and
column height moments. We found that this model has a limit distribution 
of the same type than that of the model with diagonal moments.
We first analysed the limit distribution of the column height
moments for the model of staircase polygons counted by {\it total} perimeter.
An exact analysis of the first few moment generating functions for $M=2$ and 
$M=3$, using the functional equation \eqref{form:anis}, is consistent with 
the assumption that the limit distributions of both models are of the same type. 
We also performed a Monte-Carlo analysis on self-avoiding polygons for the 
corresponding two types of vertical layer moments, where we sampled polygons
w.r.t.~half-perimeter $n_0$. We found that the moment 
ratios $m^{(a)}_{2,n_0}(2)/m^{(a)}_{2,n_0}(1)^2$ and $m^{(b)}_{2,n_0}(2)/
m^{(b)}_{2,n_0}(1)^2$ for vertical layer moments yield (within numerical 
accuracy) the same values than the corresponding values for diagonal 
layer moments. This suggests that the limit distributions are of the same type,
being however different from those of staircase polygons.

\section{Related models}

We consider some classes of directed square lattice random walks. These 
have been analysed in \cite{NT03} by a generating function approach, 
mainly according to their area laws. We extend this analysis by providing 
moment recurrences for the laws of counting parameters of generalised 
area. A main ingredient in our approach is the method of dominant balance.

All walks start at the origin. The only allowed steps are forward unit steps 
along the positive and negative diagonals. Such walks are called 
{\it Bernoulli random walks}. If the walk does not step below the 
horizontal axis, it is called a {\it meander}. A {\it Dyck path} is a 
meander terminating in the horizontal axis. A {\it bilateral Dyck path} 
is a Bernoulli random walk terminating in the horizontal axis. 
These walk models are discrete counterparts of Brownian motion, meanders, 
excursion and bridges. Corresponding convergence results appear e.g. in 
\cite{K76, A92, DM05}.

For a Bernoulli random walk $b$, let its length be the number of its steps 
$n(b)$. We are interested in the $k$-th moments of (the absolute value of) 
height, defined by $n_k(b)=\sum_{s\in w} |h(s)|^k$, where $h(s)$ is the 
height of the walk at position $s$, with $s=0,1,\ldots,n(b)$. 
Define the weight $w_b(\boldsymbol u)$ of a Bernoulli random walk $b$ by
\begin{equation*}
w_b(\boldsymbol u)=u_0^{n(b)}\cdot u_1^{n_1(b)}\cdot\ldots\cdot u_M^{n_M(b)},
\end{equation*}
and let $G^{(r)}(\boldsymbol u)=\sum_{b\in\cal B} w_b(\boldsymbol u)$ denote 
the generating function of the class $\cal B$ of Bernoulli random walks. 
Likewise, define generating functions for the other classes of random walks
by restricting the summation to the corresponding subclasses of Bernoulli random walks.

\begin{theorem}[\cite{NT03}]
Let $G^{(d)}(\boldsymbol u)$, $G^{(b)}(\boldsymbol u)$, $G^{(m)}(\boldsymbol u)$ and 
$G^{(r)}(\boldsymbol u)$ denote the generating functions of Dyck paths, bilateral 
Dyck paths, meanders, and Bernoulli random walks. The following functional equations are satisfied.
\begin{equation*}
\begin{split}
G^{(d)}(\boldsymbol u)&=\frac{1}
{1-u_0^2u_1\cdot\ldots\cdot u_MG^{(d)}(\boldsymbol v(\boldsymbol u))},\\
G^{(b)}(\boldsymbol u)&=\frac{1}
{1-2u_0^2u_1\cdot\ldots\cdot u_MG^{(d)}(\boldsymbol v(\boldsymbol u))},\\
G^{(m)}(\boldsymbol u)&=G^{(d)}(\boldsymbol u)(1+u_0\cdot\ldots\cdot u_M 
G^{(m)}(\boldsymbol v(\boldsymbol u)),\\
G^{(r)}(\boldsymbol u)&=G^{(b)}(\boldsymbol u)(1+2u_0\cdot\ldots\cdot u_M 
G^{(m)}(\boldsymbol v(\boldsymbol u)),
\end{split}
\end{equation*}
where the functions $v_k({\boldsymbol u})$ are given by
\begin{equation*}
v_k({\boldsymbol u})=\prod_{l=k}^M u_l^{\binom{l}{k}} \qquad (k=0,1,\ldots,M).
\end{equation*}
\qed
\end{theorem}

A proof of these formulae has been indicated in \cite{NT03}. The underlying 
combinatorial constructions are as follows. Dyck paths are ordered 
sequences of arches, where an arch is a Dyck path, which does not 
touch the horizontal line, except for its start point and its end point. 
Bilateral Dyck paths are ordered sequences of positive or negative arches. 
Meanders are either Dyck paths or Dyck paths, followed by a meander 
with an additional base layer attached. Bernoulli random walks are 
either bilateral random walks, or bilateral random walks followed 
by a positive or negative meander, with an additionally attached 
base layer. These constructions translate immediately into the above 
functional equations, where the same techniques as in the proof of 
Theorem \ref{theo:feq} are used. See also \cite{R05} for Dyck paths.

\vspace{2ex}

Singularity analysis of the moment generating functions can be done in analogy to
staircase polygons. Bounds on exponents can be proved by induction, using the 
corresponding functional equations. Recursions for coefficients can then be obtained 
by applying the method of dominant balance. We have the following lemma.

\begin{lemma}\label{theo:Puiwalk}
All generating functions $g^{(\cdot)}_{\boldsymbol  k}(u_0)$ are algebraic, where 
$(\cdot)\in \{(d),(b),(m),(r)\}$. They are analytic for $|u_0|\le u_c=1/2$, except at $u_0=\pm u_c$, 
with Puiseux expansions about $u_0=u_c$ of the form

\begin{equation*}
g^{(\cdot)}_{\boldsymbol k}(u_0) = \sum_{l=0}^\infty f^{(\cdot)}_{\boldsymbol  k,l}(u_c-u_0)^{l/2-
\gamma^{(\cdot)}_{\boldsymbol k}}.
\end{equation*}
The exponents $\gamma^{(\cdot)}_{\boldsymbol k}$ are given by
\begin{equation*}
\gamma^{(d)}_{\boldsymbol k}=-\frac{1}{2}+\sum_{i=1}^M\left(1+\frac{i}{2}\right) k_i, \qquad
\gamma^{(b)}_{\boldsymbol k}=\gamma^{(m)}_{\boldsymbol k}=\gamma^{(d)}_{\boldsymbol k}+1, \qquad
\gamma^{(r)}_{\boldsymbol k}=\gamma^{(d)}_{\boldsymbol k}+\frac{3}{2}.
\end{equation*}
The leading coefficients $f^{(\cdot)}_{\boldsymbol k,0}=f^{(\cdot)}_{\boldsymbol k}$ are, 
for ${\boldsymbol k}\ne {\bf 0}$, determined by the recursions
\begin{equation*}
\begin{split}
&\gamma^{(d)}_{{\boldsymbol k}-{\boldsymbol e}_1}
f^{(d)}_{{\boldsymbol k}-{\boldsymbol e}_1}+2\sum_{i=1}^{M-1}(i+1)( k_i+1)
f^{(d)}_{{\boldsymbol k}-{\boldsymbol e}_{i+1}+{\boldsymbol e}_i}+\sum_{
{\bf 0}\le{\boldsymbol l}\le{\boldsymbol k}}f^{(d)}_{\boldsymbol l}
f^{(d)}_{{\boldsymbol k}-{\boldsymbol l}}=0,\\
&\gamma^{(b)}_{{\boldsymbol k}-{\boldsymbol e}_1}
f^{(b)}_{{\boldsymbol k}-{\boldsymbol e}_1}+2\sum_{i=1}^{M-1}(i+1)( k_i+1)
f^{(b)}_{{\boldsymbol k}-{\boldsymbol e}_{i+1}+{\boldsymbol e}_i}-8\sum_{
{\bf 0}\le{\boldsymbol l}\le{\boldsymbol k}}f^{(b)}_{\boldsymbol l}
f^{(b)}_{{\boldsymbol k}-{\boldsymbol l}}=0,\\
&\gamma^{(m)}_{{\boldsymbol k}-{\boldsymbol e}_1}
f^{(m)}_{{\boldsymbol k}-{\boldsymbol e}_1}+2\sum_{i=1}^{M-1}(i+1)( k_i+1)
f^{(m)}_{{\boldsymbol k}-{\boldsymbol e}_{i+1}+{\boldsymbol e}_i}+\sum_{
{\bf 0}\le{\boldsymbol l}\le{\boldsymbol k}}f^{(m)}_{\boldsymbol l}
f^{(d)}_{{\boldsymbol k}-{\boldsymbol l}}=0,\\
&f^{(r)}_{\boldsymbol k}=\sum_{
{\bf 0}\le{\boldsymbol l}\le{\boldsymbol k}}f^{(b)}_{\boldsymbol l}
f^{(m)}_{{\boldsymbol k}-{\boldsymbol l}},
\end{split}
\end{equation*}
with boundary conditions $f^{(d)}_{\boldsymbol 0}=-4$, $f^{(b)}_{\boldsymbol 0}=1/2$, 
$f^{(m)}_{\boldsymbol 0}=1$, $f^{(r)}_{\boldsymbol 0}=1/2$, 
and $f^{(\cdot)}_{\boldsymbol k}=0$ if $ k_j<0$ for some $1\le j\le M$. The 
coefficients $f^{(\cdot)}_{\boldsymbol k}$ are strictly 
positive for ${\boldsymbol k}\ne {\bf 0}$. 
\qed
\end{lemma}

\noindent {\bf Remark.} The relations between the coefficients for Dyck paths, 
bilateral Dyck paths, meanders and Bernoulli random walks have discussed in 
\cite{NT03} in the case $M=1$, and in \cite{NT03b} partly in the case
$M=2$. The recurrence for Dyck paths for general $M$ is given in \cite{R05}.

\vspace{2ex}

The above recursions can also be phrased in terms of the corresponding 
coefficient generating functions $F^{(\cdot)}_0(\boldsymbol \epsilon)=
\sum_{\boldsymbol k} (-1)^{|\boldsymbol k|} f^{(\cdot)}_{\boldsymbol k} 
\boldsymbol \epsilon^{\boldsymbol k}$, which are (candidates for) 
scaling functions. They are given by
\begin{equation} \label{form:ssfun}
\begin{split}
&\epsilon_1\left(\frac{1}{2}F^{(d)}_0(\boldsymbol \epsilon) -
\sum_{i=1}^M \left(1+\frac{i}{2}\right)\epsilon_i 
\frac{\partial F^{(d)}_0}{\partial \epsilon_i}(\boldsymbol\epsilon)\right)
+2\sum_{i=1}^{M-1}(i+1)\epsilon_{i+1}\frac{\partial F^{(d)}_0}{\partial \epsilon_i}(\boldsymbol\epsilon)
+F^{(d)}_0(\boldsymbol\epsilon)^2 =16,\\
&F^{(b)}_0(\boldsymbol \epsilon)=-2\left(F^{(d)}_0(\boldsymbol \epsilon)\right)^{-1},\\ 
&
\epsilon_1\left(\frac{1}{2}F^{(m)}_0(\boldsymbol \epsilon) +
\sum_{i=1}^M \left(1+\frac{i}{2}\right)\epsilon_i 
\frac{\partial F^{(m)}_0}{\partial \epsilon_i}(\boldsymbol\epsilon)\right)
+2\sum_{i=1}^{M-1}(i+1)\epsilon_{i+1}\frac{\partial F^{(m)}_0}{\partial \epsilon_i}(\boldsymbol\epsilon)+\\
&+F^{(m)}_0(\boldsymbol\epsilon)F^{(d)}_0(\boldsymbol\epsilon)=4,\\
&F^{(r)}_0(\boldsymbol \epsilon)=F^{(b)}_0(\boldsymbol \epsilon)F^{(m)}_0(\boldsymbol \epsilon).
\end{split}
\end{equation}

As in section \ref{limdist}, one can now proceed in defining random variables 
and prove, after appropriate normalisation, the existence and uniqueness of a 
limit distribution. For Dyck paths and $k$-th height moments, the limit 
distributions coincide with those of $k$-th excursion moments, i.e., the 
integral $\int_0^1 e^k(t)\,{\rm d}t$ over the $k$-th power of the standard 
Brownian excursion of duration 1. In view of the bijection described in the 
introduction, this result is not unexpected. For meanders, bilateral Dyck 
paths and Bernoulli random walks, we have convergence to $k$-th moments of 
meanders, of the absolute value of Brownian bridges and Brownian motion. 
In fact, convergence results for underlying stochastic processes have been 
obtained previously \cite{K76,A92,DM05}. Note that, apart from moment 
convergence, the above method also yields moment recurrences and corrections 
to the asymptotic behaviour, which cannot easily be obtained following the 
stochastic approach, see \cite{R05} for the case of Dyck paths. Consistency 
with the stochastic description has been demonstrated for $k=1$, by deriving 
explicit expressions for the scaling functions \cite{NT03}. Also, the 
second and the last equation of \eqref{form:ssfun} have a stochastic 
counterpart, see \cite[Thm.~2]{NT03} and \cite[Thm.~7]{NT03}. 
We leave further details to the interested reader.

\section{Conclusion}

We analysed the model of staircase polygons, counted
by perimeter and $k$-th diagonal length moments, and by width and $k$-th 
column height moments. The model may serve as a toy model of 
vesicle collapse in $k+1$ dimensions. For example if $k=2$, staircase 
polygons may be converted into a three-dimensional object (vesicle) by replacing 
each column of height $n$ by an $n\times n$ slice extending into the third 
dimension. The second column height moment is then the volume of the vesicle.

We derived limit distributions for the diagonal length moments in the limit of 
large perimeter. These results may be viewed as a special case of limit 
distributions corresponding to $q$-functional equations with a square-root 
singularity as dominant singularity of the perimeter generating function \cite{R05}. 
Distributions of the same type appear for the column height moments in the limit
of large width. They also seem to appear for column height moments in the limit
of large perimeter, but a proof is an open problem.

We presented corresponding results for related, but different problems 
of directed walks such as meanders and bridges. Variants of these models, together 
with a contact activity, may be useful for questions of polymer adsorption 
\cite{J00}. Also, the methods used in this paper can be applied in the analysis of
parameters on trees, such as the Wiener index \cite{J03} and 
left and right pathlengths \cite{J04}, see also \cite{BMJ05}.

It is perhaps surprising that staircase polygons and self-avoiding
polygons might obey the same area law, but that their higher layer 
moments seem to be distributed differently. Apparently, both 
models do {\em not} share the same universality class. Understanding 
self-avoiding polygons remains a challenge for the future. 

\section*{Acknowledgements}

The author thanks an anonymous referee for a number of suggestions improving
the manuscript.

\end{document}